\documentclass[a4paper,12pt]{article}
\usepackage[left=2.5cm,bottom=3cm,right=2.5cm,top=3cm]{geometry}
 \usepackage{mathtools}  
\usepackage{makecell}   %
\usepackage{bigints}
\usepackage{slashed}
\usepackage{overpic,youngtab}
\usepackage{bbm}
\usepackage[latin,english]{babel}
\usepackage{amsmath}
\usepackage{amssymb}
\usepackage{epstopdf}
\usepackage{graphics,psfrag,rotating}
\usepackage{mathabx}
\usepackage{graphicx}
\usepackage{dcolumn}
\usepackage{float}
\usepackage{pdflscape}
\usepackage{array}
\usepackage{booktabs}
\usepackage{amscd}
\usepackage{mathtools}
\usepackage{fancybox}
\usepackage{fix-cm}
\usepackage[colorlinks=true,citecolor=blue,,linktocpage=true,linkcolor=blue,urlcolor=black]{hyperref}
\usepackage{tikz}
\usetikzlibrary{matrix}
\usetikzlibrary{positioning}
\usetikzlibrary{arrows}
\usepackage{titlesec}
\usepackage{abstract}
\newcolumntype{M}[1]{>{\centering\arraybackslash}m{#1}}
\newcolumntype{N}{@{}m{0pt}@{}}

\def\be{\begin{equation}}
\def\ee{\end{equation}}
\def\bea{\begin{eqnarray}}
\def\eea{\end{eqnarray}}
\def\pd{\partial}
\def\a{\alpha}
\def\b{\beta}
\def\g{\gamma}
\def\d{\delta}
\def\m{\mu}
\def\n{\nu}

\def\r{\rho}

\def\s{\sigma}

\def\e{\epsilon}
\def\bi{\begin{itemize}}
\def\ei{\end{itemize}}

\newcommand{\email}[1]{\href{mailto:#1}{\tt #1}}

\usepackage{epsfig,amsmath,amssymb}
\usepackage[utf8]{inputenc}
\usepackage[colorlinks=true,citecolor=blue,,linktocpage=true,linkcolor=blue,urlcolor=black]{hyperref}
\usepackage{mathtools,slashed}
\usepackage{array}
\usepackage{xcolor}
\usepackage{caption}
\usepackage{subcaption}
\usepackage{overpic,youngtab}
\usepackage[latin,english]{babel}
\usepackage{amsmath}
\usepackage{amssymb}
\usepackage{epstopdf}
\usepackage{graphics,psfrag,rotating}
\usepackage{mathabx}
\usepackage{graphicx}
\usepackage{dcolumn}
\usepackage{float}
\usepackage{pdflscape}
\usepackage{array}
\usepackage{booktabs}
\usepackage{amscd}
\usepackage{mathtools}
\usepackage{fancybox}
\usepackage{fix-cm}
\usepackage[colorlinks=true,citecolor=blue,,linktocpage=true,linkcolor=blue,urlcolor=black]{hyperref}
\usepackage{tikz}
\usetikzlibrary{matrix}
\usetikzlibrary{positioning}
\usetikzlibrary{arrows}

\renewenvironment{itemize}
  {\begin{list}%
     {}%
     {\setlength{\topsep}{0pt}%
      \setlength{\partopsep}{0pt}%
      \setlength{\itemsep}{-4pt}%
      \setlength{\labelsep}{5pt}%
      \setlength{\itemindent}{0pt}%
     }%
  }%
  {\end{list}}%

\makeatletter
\renewcommand{\section}{\setcounter{equation}{0}\@startsection
 {section}%
 {1}%
 {0pt}%
 {-1\baselineskip}%
 {0.4\baselineskip}%
 {\bfseries\large}}%
\renewcommand{\subsection}{\@startsection
 {subsection}%
 {2}%
 {0pt}%
 {-0.75\baselineskip}%
 {0.2\baselineskip}%
 {\bfseries}}%
\renewcommand{\subsubsection}{\@startsection
 {subsubsection}%
 {3}%
 {0pt}%
 {-0.5\baselineskip}%
 {0.1\baselineskip}%
 {\bfseries}}%
\makeatother

\DeclareMathAlphabet{\mathpzc}{OT1}{pzc}{m}{it}

\usepackage{tikz}
\usetikzlibrary{patterns,snakes}
\usetikzlibrary{shapes.misc}
\usetikzlibrary{decorations.markings,decorations.pathmorphing}
\usetikzlibrary{calc}
\tikzstyle{spring}=[line width=0.8,black,snake=coil,segment amplitude=4.25,segment length=4.75,line cap=round]

\def\be{\begin{equation}}
\def\ee{\end{equation}}
\def\a{\alpha}
\def\b{\beta}
\def\d{\delta}

\def\g5{\gamma_{5}}

\def\e{\eta}

\def\m{\mu}
\def\n{\nu}
\def\r{\rho}
\def\s{\sigma}

\def\id3k{\int\!\! \dfrac{d^3\!\vec{k}}{(2\pi)^3 }}
\def\idp{\int\!\! \dfrac{d^4\!p}{(2\pi)^4}}

\def\dpi{\dfrac{d^4\!p_j}{(2\pi)^4}}

\def\idq{\int\!\! \dfrac{d^4\!q}{(2\pi)^4} \,\,}

\def\idx{\int\!\! d^4\!x}

\def\idx{\int d^{4}\!x}

\newcommand{\beann}{\begin{eqnarray*}}
\newcommand{\eeann}{\end{eqnarray*}}
\newcommand{\ba}{\begin{array}}
\newcommand{\ea}{\end{array}}

 \def\g {\gamma}

\def\bsigma {\bar{\sigma}}

\begin{document}

		\vspace*{-1cm}
		\phantom{hep-ph/***}
		{\flushleft
			{{FTUAM-}}
			\hfill{{ IFT-UAM/CSIC-25-153}}}
		\vskip 1.5cm
		\begin{center}
		{\LARGE\bfseries On the Weyl  anomaly for chiral fermions.}\\
		
		
			\vskip .3cm
		
		\end{center}
		\vskip 0.5  cm
		\begin{center}
			{\large Enrique \'Alvarez$^{(1)}$, Luis \'Alvarez-Gaum\'e$^{(2)}$,\\ Jes\'us Anero $^{(1)}$ and Carmelo P Mart\'{\i}n$^{(3)}$  }
			\\
			\vskip .7cm
			{
				1.-Departamento de F\'isica Te\'orica and Instituto de F\'{\i}sica Te\'orica,
				IFT-UAM/CSIC,\\
				Universidad Aut\'onoma de Madrid, Cantoblanco, 28049, Madrid, Spain\\
				2.-Simons Center for Geometry and Physics,
State University of New York Stony Brook,
NY 11794 3636, USA\\
Physics Department  (Emeritus) \\
CERN, CH-1211 Geneva 23\\
3.- Universidad Complutense de Madrid (UCM), Departamento de F\'{\i}sica Te\'orica and
IPARCOS, Facultad de Ciencias F\'{\i}sicas, 28040 Madrid, Spain

				\vskip .1cm

				\vskip .5cm
				\begin{minipage}[l]{.9\textwidth}
					\begin{center}
						\textit{E-mail:}
						\email{enrique.alvarez@uam.es,jesusanero@gmail.com, carmelop@fis.ucm.es,lalvarezgaume@scgp.stonybrook.edu}

					\end{center}
				\end{minipage}
			}
		\end{center}
	\thispagestyle{empty}
	
\begin{abstract}\vspace{-1em}
	\noindent
We compute the parity-odd part of the Weyl anomaly for chiral fermions in a background gravitational field. We start from a manifestly real form of the Lagrangian (that is, not only real up to a total derivative), and we regularize it by means of Pauli-Villars fermions. All parity-odd terms in the anomaly cancel in the {\em integrand}, so that the result of the anomaly is necessarily parity-even.
\end{abstract}
\newpage
\tableofcontents
	\thispagestyle{empty}
\flushbottom

\newpage
\setcounter{page}{1}
\section{Introduction}
There has recently been some interest in the Weyl (conformal) anomaly of a chiral  fermion on a background  gravitational field with metric $g_{\m\n}$. This was motivated by a paper by Bonora, Giaccari and Lima de Souza (cf. the first paper in \cite{Bonora}) claiming that there is a parity-odd term in the trace anomaly proportional to the Pontryagin index.  Symbollically
\be
\langle g^{\m\n} T_{\m\n}\rangle\sim i \int d(vol)\,R _{\m\n\r\s}\eta^{\r\s \a\b} R_{\a\b}^{~~~\m\n}
\ee
with an {\em imaginary} coefficient. Here  $R _{\m\n\r\s}$ are the components of Riemann's tensor and  $d(vol)$ is the usual  volume element
\be
d(vol)\equiv {1\over 4!}\,\eta_{\m\n\r\s}dx^{\m\n\r\s}\equiv {1\over 4!}\,\sqrt{|g|} \e_{\m\n\r\s}dx^\m\wedge dx^\n\wedge dx^\r\wedge dx^\s
\ee
In addition, in \cite{Bonora} it was claimed that this contribution was suggested earlier by Christensen and Duff \cite{Christensen}. Actually, in \cite{Christensen} the conformal anomaly corresponding to an euclidean  field transforming with  the $(A,B)$ representation of the $SO(4)$ group is determined in terms of the Schwinger-DeWitt coefficients as 
\be
T^\m_\m={1\over 2}\,(-1)^{2A+2B} \left(b_4(A,B)+b_4(B,A)\right)
\ee
Although both $b_2(1/2,0)$ and $b_2(0,1/2)$ contain pieces proportional to Pontryagin's term, they cancel in the sum, which is indeed parity even.

\par
This term would violate unitarity and point to a fatal inconsistency in the physics of such chiral fermions. The main reason is that the energy momentum tensor is the coefficient of the response of the action  to an arbitrary (but real) variation of the metric
\be
\d S\equiv \int d(vol) T_{\m\n} \d g^{\m\n}
\ee
If the trace of the energy momentum tensor is not real, then the action itself is also not real, which violates unitarity among other things.
\par

This result however disagrees with the computations of some authors \cite{Abdallah,Bastianelli,Christensen,Frob, Gipson,Larue,Stanev}, but  some other works  like \cite{Liu} support  it.  A useful summary of the literature discussing the proposal can be found in the book \cite{Book}.\\

\par
To be precise, in reference  \cite{Bonora}  it is argued that the total conformal anomaly of a chiral fermion is given by
\be
T_\mu{}^\mu= \frac{1}{180\times16\pi^{2}}\left(\frac {11}{4} \,  E_4-\frac {9}{2}\, W^2 +i \frac{15}{4}\, P\right)\
\ee
here $E_4$ is the integrand of the Euler characteristic, a topological invariant given by
\bea
\chi(M)&&\equiv {1\over 32\pi^2}\int \e_{abcd} R^{ab}\wedge R^{cd}=\int d(vol)E_4= {1\over 2}{1\over (4\pi)^2}\int d(vol) R^* R^*=\nonumber\\
&&={1\over 32\pi^2} \int d(vol)\left(R_{\m\n\r\s}^2 -4 R_{\m\n}^2+ R^2\right)
\eea
and $W^2$ is the square of Weyl's tensor, which  can be characterized by being  invariant under conformal transformations
\be
g_{\m\n}\rightarrow \Omega^2 g_{\m\n}
\ee
namely
\be
W^\m\,_{\n\r\s}\rightarrow W^\m\,_{\n\r\s}
\ee
It follows that $\sqrt{|g|} W_{\m\n\r\s} W^{\m\n\r\s}$ is a conformal singlet in $n=4$ dimensions.

\be
W^2\equiv W_{\m\n\r\s}W^{\m\n\r\s}=R_{\m\n\r\s}^2-2 R_{\m\n}^2+{1\over 3} R^2
\ee
and $P_1(M)$ is the Pontryagin density (another topological invariant \cite{Eguchi})
\be
P_1(M)\equiv -{1\over 8\pi^2}\int \text{tr} R\wedge R={1\over (4\pi)^2} \int d(vol) R_{\m\n\r\s}^* R^{\m\n\r\s}
\ee
The dual Riemann tensor is given by
\be
R^*_{\m\n\a\b}\equiv  {1\over 2}\eta_{\m\n\r\s}R^{\r\s}\,_{\a\b}
\ee
Our aim in this paper is to  compute this anomaly using a different, and in our opinion, safer procedure \cite{Vilenkin,Okuyama,Godazgar}. We will start from a real Lagrangian and regulate the theory using Pauli-Villars fermions in an attempt to avoid the ambiguities of dimensional regularization in the presence of  $\g_5$.
\par
Our main interest will be in the parity-odd part of the anomaly (the coefficient of the Pontryagin invariant) which  is where there is disagreement in the literature.

Let us now be more precise about the definition of the problem.

Let $\omega_{\mu a b}(x)$ be the Lorentz (spin) connection defined in terms of the Levi-Civita one by
\begin{equation*}
\omega_{\mu a b}(x)= -e^{\lambda}_{b}(x)\partial_\mu e_{\lambda a}(x)+e_{\lambda a}(x)\Gamma^{\lambda}_{\rho \mu}e^{\rho}_{b}(x).
\end{equation*}
Then, the classical action of the model reads
\begin{equation*}
S=\cfrac{1}{2}\,\idx\,e(x)\,e^{\mu}_{a}(x)\,[i\bar{\xi}_{\dot{\alpha}}(x)\bsigma^{\dot{\alpha}\alpha a}({\cal D}_{\mu}\xi)_{\alpha}(x)-i
({\cal D}_\mu\bar{\xi})_{\dot{\alpha}}(x)\bsigma^{\dot{\alpha}\alpha a}\xi_{\alpha}(x)],
\end{equation*}
where
\begin{equation*}
({\cal D}_{\mu}\xi)_{\alpha}(x)=\partial_\mu\xi_{\alpha}(x)-\cfrac{i}{2}\omega_{\mu a b}(x)\,(\sigma^{a b}\xi)_{\alpha}(x),\quad
({\cal D}\bar{\xi})_{\dot{\alpha}}(x)=\partial_{\mu}\bar{\xi}_{\dot{\alpha}}(x)+\cfrac{i}{2}(\bar{\xi}(x)\bar{\sigma}^{ab})_{\dot{\alpha}}\,\omega_{\mu a b}(x).
\end{equation*}
We shall denote spinorial indices by the first letters of the Greek alphabet ($\a,\dot{\a},\b,\dot{\b}.\ldots=1,2 $) and reserve the middle letters of the Greek alphabet for four-dimensional vector indices ($\m,\n,\ldots=0,\ldots, 3 $)
Our convention \cite{Dreiner} for the Pauli matrices is as follows
\bea
&\s^0=\overline{\s}^0\equiv \begin{pmatrix} 1&0\\0&1\end{pmatrix}\nonumber\\
&\s^1=-\overline{\s}^1\equiv \begin{pmatrix} 0&1\\1&0\end{pmatrix}\nonumber\\
&\s^2=-\overline{\s}^2\equiv \begin{pmatrix} 0& -i\\i&0\end{pmatrix}\nonumber\\
&\s^3=-\overline{\s}^3\equiv \begin{pmatrix} 1&0\\0&-1\end{pmatrix}
\eea
There is another notation frequently used representing chiral spinors as left and right ones, $\psi_L$ or $\psi_R$. The relationship with ours is
\bea
&\psi_L\leftrightarrow \xi_\a\nonumber\\
&\psi_R\leftrightarrow \overline{\xi}_{\dot{\a}}
\eea

In is important to remark that the action is real without total derivatives. This fact will be relevant later on.
\par
In constructing the quantum theory for the  fermions in perturbation theory around Minkowski spacetime, we shall express the metric, $g_{\mu\nu}$, the vierbein $e^a_\mu$, the inverse vierbein, $e^{\mu}_a$, and the spin connection, $\omega_{a, bc}$ as follows (the perturbative parameter is related to Newton's constant by $\kappa^2\equiv 8\pi G$)
\begin{equation*}
\begin{array}{l}
{g_{\mu\nu}=\eta_{\mu\nu}+\kappa h_{\mu\nu},}\\[4pt]
{e^a_{\mu}=\delta^a_\mu+\kappa{\cal E}^{(1)\,a}_{\mu}+\kappa^2 {\cal E}^{(2)\,a}_{\mu}+...}\\[4pt]
{e^\mu_{a}=\delta^\mu_a+\kappa{\cal E}^{(1)\,\mu}_{a}+\kappa^2 {\cal E}^{(2)\,\mu}_{a}+...}\\[4pt]
{\omega_{a, bc}=e^{\mu}_a\omega_{\mu bc}=\kappa \Omega^{(1)}_{a, bc}+\kappa^2 \Omega^{(2)}_{a, bc}+....,}
\end{array}
\end{equation*}
in the above the terms involving the external gravitons become
\begin{equation*}
\begin{array}{l}
{{\cal E}^{(1)\,a}_{\mu}=\cfrac{1}{2}h_a^{\mu},\quad {\cal E}^{(2)\,a}_{\mu}=-\frac{1}{8}\,h^{a\rho}h_{\rho\mu},\quad
{\cal E}^{(1)\,\mu}_{a}=-\cfrac{1}{2}h^\mu_a,\quad {\cal E}^{(2)\,\mu}_{a}=\frac{3}{8}\,h^{\mu\rho}h_{\rho a}}\\[4pt]
{\Omega^{(1)}_{a, bc}=-\frac{1}{2}\,(\partial_b h_{ca}-\partial_c h_ba)}\\[4pt]
{\Omega^{(2)}_{a, bc}=\frac{1}{4}h^{\mu}_a(\partial_b h_{c\mu}-\partial_c h_{b\mu})-\frac{1}{8}(h^{\rho}_b\partial_a h_{c\rho}-
h^{\rho}_c\partial_a h_{b\rho})+\frac{1}{4}(h^{\rho}_b\partial_\rho h_{c a}-h^{\rho}_c\partial_\rho h_{b a})-\frac{1}{4}
(h^{\rho}_b\partial_c h_{\rho a}-h^{\rho}_c\partial_b h_{\rho a}).}
\end{array}
\end{equation*}
The linearized conformal transformation reads
\be
\d_\theta h_{\m\n}=2\left({\theta(x)\over \kappa} \eta_{\m\n}+h_{\m\n}\right)
\ee

The previous expansions in powers of $\kappa$, leads to the following expansion of the action $S$:
\begin{equation*}
\begin{array}{l}
{S=S_{0}+S_{int}}\\[4pt]
{S_{0}=\cfrac{1}{2}\,\idx\,\,[i\bar{\xi}_{\dot{\alpha}}(x)\bsigma^{\dot{\alpha}\alpha\mu}(\partial_{\mu}\xi)_{\alpha}(x)-i
(\partial_\mu\bar{\xi})_{\dot{\alpha}}(x)\bsigma^{\dot{\alpha}\alpha \mu}\xi_{\alpha}(x)],}\\[4pt]
{S_{int}=\sum\limits_n\,\kappa^n S^{(n)}.}
\end{array}
\end{equation*}
In both $S_0$ and $S_{int}$ all the indices are flat. $S_0$ is the free action in Minkowski spacetime and $S_{int}$ carry the interaction vertices between the graviton field $h_{\mu\nu}$ and the fermions. $S^{(n)}$ is quadratic on the fermion fields and involves $n$ powers of $h_{\mu\nu}$ and/or its derivatives.

Let ${\cal Z}[h_{\mu\nu}]$ be partition function of the model is defined by perturbation theory around the Minkowski background
\begin{equation*}
\begin{array}{l}
{{\cal Z}[h_{\mu\nu}]=\cfrac{1}{{\cal N}}\int \mathcal{D}\xi_{\alpha}\mathcal{D}\bar{\xi}_{\dot{\alpha}}\;e^{i S}=\langle\,e^{iS_{int}}\,\rangle_0}\\[4pt]
{{\cal N}=\int \mathcal{D}\xi_{\alpha}\mathcal{D}\bar{\xi}_{\dot{\alpha}}\;e^{i S_{0}}.}
\end{array}
\end{equation*}
$\langle {\cal O}\rangle_{0}$ denotes the average of ${\cal O}$  with regard to the free action $S_{0}$:
\begin{equation*}
\langle {\cal O}\rangle_{0}=\cfrac{1}{{\cal N}}\int \mathcal{D}\xi_{\alpha}\mathcal{D}\bar{\xi}_{\dot{\alpha}}\,{\cal O}\,e^{i S_0}
\end{equation*}

Then, the quantum action, ${\cal W}[h_{\mu\nu}]$, of the model is defined to be
\begin{equation}
{\cal W}[h_{\mu\nu}]= -i\, {\rm Ln} {\cal Z}[h_{\mu\nu}]=-i\sum_n\,\cfrac{i^n}{n!}\,\langle (S_{int})^n\rangle_{0}^{(connected)},
\label{Qaction}
\end{equation}
The superscript "connected" signals that only connected contractions, as given by Wick's theorem, are to be kept.
\subsection{Pauli-Villars fields and the regularized quantum action.}
As it stands, ${\cal W}[h_{\mu\nu}]$, (\ref{Qaction}) is an ill-defined  due to ultraviolet (UV)  divergences. We shall associate to this ill-definite quantity a regularized --and, therefore, well-defined mathematically-- expression. This will be done by introducing several massive Pauli-Villars (PV) bispinor fields:  $\xi^{(i)}_{\alpha}$, $\bar{\xi}^{(i)}_{\dot{\alpha}}$, $i=1...f$, with classical actions
\begin{equation*}
\begin{array}{l}
{S^{(i)}=}\\[4pt]
{\frac{1}{2}\,\idx\,e(x)\,e^{\mu}_{a}(x)\,[i\bar{\xi}^{(i)}_{\dot{\alpha}}(x)\bsigma^{\dot{\alpha}\alpha a}({\cal D}_{\mu}\xi^{(i)})_{\alpha}(x)-i
({\cal D}\bar{\xi}^{(i)})_{\dot{\alpha}}(x)\bsigma^{\dot{\alpha}\alpha a}\xi^{(i)}_{\alpha}(x)]}\\[4pt]
{-\frac{m_i}{2}\,\idx\,e(x)\,\,[\xi^{(i)\,\alpha}(x) \xi^{(i)}_{\alpha}(x)
+\bar{\xi}^{(i)}_{\dot{\alpha}}(x)\bar{\xi}^{(i)\,\dot{\alpha}}(x)].}
\end{array}
\end{equation*}
The usual PV regulators (as explained, for example in Faddeev and Slavnov's book \cite{Faddeev} page 131, or in  Itzykson and Zuber's \cite{Itzykson} page 411) consist in replacing some fermionic gaussian integral
\be 
\det\,\slashed{D}
\ee
by 
\be 
\det\,\slashed{D}\prod_{j=1}^f \,\left(\det\,\slashed{D}_j\right)^{c_j}
\ee
where the massive covariant derivatives are defined as
\be
\slashed{D}_j\,\equiv \slashed{D}-m_j
\ee
where the set of masses $m_j$ and constants $c_j$  are determined by some equations that will be made explicit  in what follows.
In our case the gaussian integral does not really define a determinant, but the regularization proceeds by replacing everywhere in the above formulas the determinant by the gaussian integral,

By using  the PV regulators, we define
\begin{equation}
{\cal W}^{(i)}[h_{\mu\nu}]=-i\, {\rm Ln} {\cal Z}_i [h_{\mu\nu}]=-i\,\sum_n\,\cfrac{i}{n!}\,\langle (S^{(i)}_{int})^n\rangle_{0}^{(connected)},\quad
i=1...f.
\label{WPVi}
\end{equation}
Finally, we regularize  ${\cal W}[h_{\mu\nu}]$ by replacing it with ${\cal W}^{(PV)}[h_{\mu\nu}]$, the latter defined as follows
\begin{equation}
 {\cal W}^{(PV)}[h_{\mu\nu}]=\sum\limits_{i=0}^{f}\,c_{i}\,{\cal W}^{(i)}[h_{\mu\nu}],
\label{RegW}
\end{equation}
with the understanding that $i=0$ labels the physical field: $\xi^{(0)}_{\alpha}=\xi_{\alpha}$, $\bar{\xi}^{(0)}_{\dot{\alpha}}=\bar{\xi}_{\dot{\alpha}}$,  $m_0=0$ and $c_0=1$.

The $c_{i}$'s above are the Pauli-Villars (PV) parameters and, along with Pauli-Villars masses, are to be chosen so that the loop integrals  are UV finite by power-counting at non exceptional momenta.
This regularization breaks fermion number owing to the PV mass terms, but respects diffeomorphism invariance. Any dependence on the PV masses will disappear in the physical theory.
\par
It is well known that when dealing with anomalies it is of paramount importance not to carry out any kind of simplification in the integrands of the non-regularized UV divergent integrals. So, here, we shall be careful and choose the $c_i$'s and the Pauli-Villars masses so that all the loop integrals are regularized before carrying out any manipulations in the integrands of the loop integrals.

It is not difficult to come to the conclusion that the most UV divergent  loop integrals which occur in ${\cal W}[h_{\mu\nu}]$ are
\begin{equation*}
\idq\,\cfrac{q_{\mu_1}q_{\mu_2}\cdots q_{\mu_{2n}}}{(q+P_1)^2(q+P_2)^2\cdots(q+P_n)^2},
\end{equation*}
where the $P_i$'s are  (non-exceptional) linear combinations of the external momenta. The Pauli-Villars regularization of the previous integral amounts to its replacement with
\begin{equation}
\idq\,\Big[\sum\limits_{i=0}^{f}\,c_i\,\cfrac{q_{\mu_1}q_{\mu_2}\cdots q_{\mu_{2n}}}{((q+P_1)^2-m_i^2)((q+P_2)^2-m_i^2)\cdots((q+P_n)^2-m_i^2)}\Big].
\label{inte}
\end{equation}
In  (1.17) we have added to the action fields with similar couplings to gravity as the
original fields but with equal or opposite statistics, namely, the coefficients $c_i$
can be negative.  The masses $m_j$ and coefficients $c_i$ are chosen in
such a way that the ratio is UV finite. 
\par
In fact all these integrals will be UV finite by power-counting if the $c_i$'s and $m_i$'s satisfy the Pauli-Villars conditions, which, in the case at hand, become:
\begin{equation*}
c_0=1,\quad\quad\sum\limits_{i=0}^{f}\,c_i=0,\,\sum\limits_{i=1}^{f}\,c_i\,m_i^2=0 \quad\quad\text{and}\quad \sum\limits_{i=1}^{f}\,c_i\,m_i^4=0.
\end{equation*}
The condition on $m_i^4$ has to do with the fact that the integrals in (\ref{inte}) diverge as the fourth power of the loop momentum. An explicit set that satisfies the above conditions is given by

\bea
&m_1=M;\quad m_2=M\sqrt{2};\quad m_3=M\sqrt{3}\nonumber\\
&c_1=-3;\quad c_2=3;\quad c_3=-1
\eea

But this is not all, for the mass terms of the Pauli-Villars fields also couple to gravity so that we shall have interaction vertices linear in the $m_i$'s. These vertices do not occur in the massless physical theory. Further, unlike in the massless case, the free propagators
\begin{equation}
\begin{array}{l}
{\langle\xi^{(i)}_{\alpha}(x)\xi^{(i)}_{\beta}(y)\rangle_{0}=\idp\, m_i\, \epsilon_{\beta\alpha}\,G_i(p)\,e^{-ip(x-y)},}\\[4pt]
{\langle{\bar\xi}^{(i)}_{\dot{\alpha}}(x)\bar{\xi}^{(i)}_{\dot{\beta}}(y)\rangle_{0}=\idp\, m_i\, \epsilon_{\dot{\alpha}\dot{\beta}}\,G_i(p)\,e^{-ip(x-y)},}\\[4pt]
\end{array}
\label{xixiprop}
\end{equation}
have numerators   linear in the $m_i's$.

Now, the integrals in (\ref{inte}) come from contributions where only propagators  of the type $\langle\xi^{(i)}_{\alpha}(x)\bar{\xi}^{(i)}_{\dot{\alpha}}(y)\rangle_{0}$ and only vertices which read
\begin{equation}
\idx\,H^{\mu}_{\nu}[h](x)(\bar{\xi}\bar{\sigma}^{\nu}\partial_{\mu}\xi-\partial_\mu\bar{\xi}\bar{\sigma}^{\nu}\xi)(x),
\label{partialvertex}
\end{equation}
$H^{\mu}_{\nu}[h](x)$ being a function of  $h_{\mu\nu}$, do occur. Hence,  each time a propagator of the type $\xi\bar{\xi}$ is replaced, in a integral such that (\ref{inte}) by a propagator of the kind displayed in (\ref{xixiprop}), the UV degree of divergence of the corresponding integral decreases by one unit. The same 
decrease by one unit takes place when a vertex of type (\ref{partialvertex}) is replaced by a vertex linear in $m_i$. We thus conclude that only integrals with fewer than 5 powers of $m_i$ can be UV divergent: recall that the UV degree of divergence of the integral in (\ref{inte}) is 4. By keeping track  how --through Wick contractions and vertices--  the $m_i$'s pop up in the numerator of the Feynman integrals, it is not difficult to convince oneself that the $m_i$'s always occur in even powers. To conclude,  we will meet UV divergent integrals of the type
\begin{equation*}
\idq\,\Big[\sum\limits_{i=1}^{f}\,c_i\,m_i^{2n'}\cfrac{q_{\mu_1}q_{\mu_2}\cdots q_{\mu_{2(n -n')}}}{((q+P_1)^2-m_i^2)((q+P_2)^2-m_i^2)\cdots((q+P_n)^2-m_i^2)}\Big],
\end{equation*}
with $n'=1,2$. Making these integrals UV finite does not require new Pauli-Villars conditions.

\section{No parity-odd terms in the conformal  anomaly.}

A simple way of understanding the Weyl  anomaly  stems from a Weyl transformation  in the path integral \cite{Fujikawa}. This will be enough to show that no parity-odd terms appear in the anomaly. Recall that
\begin{equation*}
\begin{array}{l}
{S[g,\xi,\bar{\xi}]=\cfrac{1}{2}\,\idx\,e(x)\,e^{\mu}_{a}(x)\,[i\bar{\xi}_{\dot{\alpha}}(x)\bsigma^{\dot{\alpha}\alpha a}({\cal D}_{\mu}\xi)_{\alpha}(x)-i
({\cal D}_\mu\bar{\xi})_{\dot{\alpha}}(x)\bsigma^{\dot{\alpha}\alpha a}\xi_{\alpha}(x)],}\\[12pt]
{S^{(i)}[g,\xi^{(i)},\bar{\xi}^{(i)}]=\cfrac{1}{2}\,\idx\,e(x)\,e^{\mu}_{a}(x)\,[i\bar{\xi}^{(i)}_{\dot{\alpha}}(x)\bsigma^{\dot{\alpha}\alpha a}({\cal D}_{\mu}\xi)^{(i)}_{\alpha}(x)-i
({\cal D}_\mu\bar{\xi}^{(i)})_{\dot{\alpha}}(x)\bsigma^{\dot{\alpha}\alpha a}\xi^{(i)}_{\alpha}(x)],}\\[12pt]
{S^{(i)}_m[g,\xi^{(i)},\bar{\xi}^{(i)}]=-\frac{m_i}{2} \idx\,e(x)\,\,[\xi^{(i)\,\alpha}(x) \xi^{(i)}_{\alpha}(x)
+\bar{\xi}^{(i)}_{\dot{\alpha}}(x)\bar{\xi}^{(i)\,\dot{\alpha}}(x)],}
\end{array}
\end{equation*}

and that in (\ref{RegW}) we defined the regularized Pauli-Villars effective action ${\cal W}^{(PV)}[g_{\mu\nu}]$ by
\begin{equation*}
{\cal W}^{(PV)}[g_{\mu\nu}]=-i\sum\limits_{i=0}^{f}\,c_i\,\ln\,{\cal Z}_i[g_{\mu\nu}]
\end{equation*}
where
\begin{equation*}
{\cal Z}_i[g_{\mu\nu}]=
\cfrac{1}{{\cal N}_i}\int \mathcal{D}\xi^{(i)}_{\alpha}\mathcal{D}\bar{\xi}^{(i)}_{\dot{\alpha}}\;
e^{i  S^{(i)}[g,\xi^{(i)},\bar{\xi}^{(i)}]+i S_m^{(i)}[g,\xi^{(i)},\bar{\xi}^{(i)}]}
\end{equation*}

Then by performing the Weyl   transformations
\begin{equation}
{g_{\mu\nu}\rightarrow \Omega^2\,g_{\mu\nu}},\quad \xi\rightarrow \Omega^{-3/2}\xi,\quad \bar{\xi}\rightarrow \Omega^{-3/2}\bar{\xi},\quad
\quad \xi^{(i)}\rightarrow \Omega^{-3/2}\xi^{(i)},\quad \bar{\xi}^{(i)}\rightarrow \Omega^{-3/2}\bar{\xi}^{(i)},
\end{equation}
we get
\begin{equation}
{\cal W}^{(PV)}[\Omega^2 g_{\mu\nu}]=-i\sum\limits_{i=0}^{f}\,c_i\,\ln\,{\cal Z}_i[\Omega^2 g_{\mu\nu}]=-i\sum\limits_{i=0}^f\,c_i\,\ln\cfrac{1}{{\cal N}_i}\int \mathcal{D}\xi^{(i)}_{\alpha}\mathcal{D}\bar{\xi}^{(i)}_{\dot{\alpha}}\;
e^{i S^{(i)}[ g,\xi^{(i)},\bar{\xi}^{(i)}]+i S_m^{(i)}[ g,\xi^{(i)},\bar{\xi}^{(i)};\Omega]},
\label{finiteform}
\end{equation}
where the transformed PV mass terms read
\begin{equation*}
{S^{(i)}_m[g,\xi^{(i)},\bar{\xi}^{(i)};\Omega]=-\frac{m_i}{2} \idx\,\Omega(x)e(x)\,\,[\xi^{(i)\,\alpha}(x) \xi^{(i)}_{\alpha}(x)
+\bar{\xi}^{(i)}_{\dot{\alpha}}(x)\bar{\xi}^{(i)\,\dot{\alpha}}(x)].}
\end{equation*}

Let us linearize around the unit transformation
\begin{equation*}
\Omega(x)=1+\theta(x),\quad 0\leq\theta(x)<< 1.
\end{equation*}
Now, the infinitesimal version of (\ref{finiteform}) reads
\begin{equation}
\delta_\theta{\cal W}^{(PV)}[g_{\mu\nu}]=\sum\limits_{i=1}^f\,c_i\,
\cfrac{1}{{\cal Z}_i[g]{\cal N}_i}\int \mathcal{D}\xi^{(i)}_{\alpha}\mathcal{D}\bar{\xi}^{(i)}_{\dot{\alpha}}\;\
(\delta_\theta S^{(i)}_{m})\;e^{i S^{(i)}[ g,\xi^{(i)},\bar{\xi}^{(i)}]+i S_m^{(i)}[ g,\xi^{(i)},\bar{\xi}^{(i)}]},
\label{AnomalyLAG}
\end{equation}
where
\begin{equation*}
\delta_\theta S^{(i)}_m=-\frac{1}{2}\,m_i\,\idx\,\theta(x)\,e(x)\,\,[\xi^{(i)\,\alpha}(x) \xi^{(i)}_{\alpha}(x)
 +\bar{\xi}^{(i)}_{\dot{\alpha}}(x)\bar{\xi}^{(i)\,\dot{\alpha}}(x)], \quad i\geq 1.
\end{equation*}

 The right hand side of (\ref{AnomalyLAG}) yields the anomaly, and at order two in the number of graviton fields $h_{\mu\nu}$ it is given by
\begin{equation}
\begin{array}{l}
{ \mathcal{A}_2=
-\frac{1}{2}\sum\limits_{i=1}^f\,c_i\,\langle \delta^{(0)}_\theta S^{(i)}_{m} S^{(i),1}_{kin}S^{(i)
,1}_{kin}\rangle_0^{(c)}+
i\,\sum\limits_{i=1}^f\,c_i\,\langle \delta^{(0)}_\theta S^{(i)}_{m}  S^{(i),2}_{kin}\rangle_0^{(c)}+
i\,\sum\limits_{i=1}^f\,c_i\,\langle \delta^{(0)}_\theta S^{(i)}_{m}  S^{(i),2}_{spin}\rangle_0^{(c)}+}\\[4pt]
{\phantom{  \mathcal{A}_2=-\frac{1}{2} }
 i\,\sum\limits_{i=1}^f\,c_i\,\langle \delta^{(1)}_\theta S^{(i)}_{m}  S^{(i),1}_{kin}\rangle_0^{(c)}+
\sum\limits_{i=1}^f\,c_i\,\langle \delta^{(2)}_\theta S^{(i)}_{m}\rangle_0^{(c)},}
\end{array}
\label{Anomalytwo}
\end{equation}
Let us recall that  the superscript $(c)$ stands for the connected contributions, the subscript $0$ denotes vaccum expectation vvalue when $h_{\mu\nu}$ is set to zero, and the Weyl variation of the PV mass terms become
\begin{equation*}
\begin{array}{l}
{\delta^{(0)}_\theta S^{(i)}_m=-\frac{1}{2}\,m_i\,\idx\,\theta(x)\,[\xi^{(i)\alpha}(x) \xi^{(i)}_{\alpha}(x)
                                 +\bar{\xi}^{(i)}_{\dot{\alpha}}(x)\bar{\xi}^{(i)\,\dot{\alpha}}(x)],}\\[4pt]
{\delta^{(1)}_\theta S^{(i)}_m=-\frac{1}{4}\,m_i\,\idx\,\theta(x)\,h^{\mu}_{\mu}(x)\,[\xi^{(i)\alpha}(x) \xi^{(i)}_{\alpha}(x)
 +\bar{\xi}^{(i)}_{\dot{\alpha}}(x)\bar{\xi}^{(i)\,\dot{\alpha}}(x)],}\\[4pt]
{\delta^{(2)}_\theta S^{(i)}_m=-\frac{1}{4}\,m_i\,\idx\,\theta(x)\,H_2(x)\,[\xi^{(i)\alpha}(x) \xi^{(i)}_{\alpha}(x)
                                 +\bar{\xi}^{(i)}_{\dot{\alpha}}(x)\bar{\xi}^{(i)\,\dot{\alpha}}(x)],}
 \end{array}
\end{equation*}
 and the kinetic and spin parts are defined as
\begin{equation*}
\begin{array}{l}
{S^{(i),1}_{kin}=\idx\,\Big\{
\frac{i}{2}\,H^{\mu}_{1,\,\nu}(x)\,[\bar{\xi}^{(i)}_{\dot{\alpha}}(x)\bsigma^{\dot{\alpha}\alpha\nu}\partial_{\mu}\xi^{(i)}_{\alpha}(x)-
\partial_\mu\bar{\xi}^{(i)}_{\dot{\alpha}}(x)\bsigma^{\dot{\alpha}\alpha \nu}\xi^{(i)}_{\alpha}(x)]}\\[4pt]
{\quad\quad\quad\quad\quad\quad\quad-\frac{m_i}{4}h^{\m}_{\mu}(x)\,[\xi^{(i)\alpha}(x) \xi^{(i)}_{\alpha}(x)
+\bar{\xi}^{(i)}_{\dot{\alpha}}(x)\bar{\xi}^{(i)\,\dot{\alpha}}(x)]\Big\},}\\[4pt]
{S^{(i),2}_{kin}=\idx\,\Big\{
\frac{i}{2}\,H^{\mu}_{2,\,\nu}(x)\,[\bar{\xi}^{(i)}_{\dot{\alpha}}(x)\bsigma^{\dot{\alpha}\alpha\nu}\partial_{\mu}\xi^{(i)}_{\alpha}(x)-
\partial_\mu\bar{\xi}^{(i)}_{\dot{\alpha}}(x)\bsigma^{\dot{\alpha}\alpha \nu}\xi^{(i)}_{\alpha}(x)]}\\[4pt]
{\quad\quad\quad\quad\quad\quad\quad-\frac{m_i}{4}H_2(x)\,[\xi^{(i)\alpha}(x) \xi^{(i)}_{\alpha}(x)
+\bar{\xi}^{(i)}_{\dot{\alpha}}(x)\bar{\xi}^{(i)\,\dot{\alpha}}(x)]\Big\},}\\[4pt]
{S^{(i),2}_{spin}=\idx\,\frac{1}{16}\,\epsilon^{\mu\nu\rho\sigma}\,h^\lambda_{\mu}(x)\partial_{\nu}h_{\rho\lambda}(x)
\bar{\xi}^{(i)}_{\dot{\alpha}}(x)\bsigma^{\dot{\alpha}\alpha}_\sigma\xi^{(i)}_\alpha(x),}\\[12pt]
{H^{\mu}_{1,\,\nu}(x)=\frac{1}{2}\,(h^{\rho}_{\rho}(x)\delta^{\mu}_{\nu}-h^{\mu}_{\nu}(x)),}\\[4pt]
{H^{\mu}_{2,\,\nu}(x)=-\frac{1}{4}h^{\rho}_{\rho}(x)h^{\mu}_{\nu}(x)+\frac{1}{8}(h^{\rho}_{\rho}(x))^2\delta^{\mu}_{\nu}-\frac{1}{4}
h^{\rho\sigma}(x)h_{\rho\sigma}(x)+\frac{3}{8}h^{\mu\rho}(x)h_{\rho \nu}(x),}\\[4pt]
{H_2(x)=\frac{1}{4} (h^{\rho}_{\rho}(x))^2-\frac{1}{2}h^{\rho\sigma}(x)h_{\rho\sigma}(x).}
\end{array}
\end{equation*}

\subsection{No parity-odd anomaly from Triangles.}
There are terms including a variation of a PV mass and two terms from the kinetic terms that would give rise to Feynman diagrams type {\em triangle}  upon Wick contractions. Let us examine them

First of all,
\begin{equation*}
-\frac{1}{2}\sum\limits_{i=1}^f\,c_i\,\langle \delta^{(0)}_\theta S^{(i)}_{m} S^{(i),1}_{kin}S^{(i),1}_{kin}\rangle_0^{(c)}=-\frac{1}{2}({\mathcal{T}}_1+ {\mathcal{T}}_2+{\mathcal{T}}_3)
\end{equation*}
This  (\ref{Anomalytwo}) yields no parity-odd anomalous contribution. In the previous equation, ${\mathcal{T}}_m$, $m=1,2,3$, is given by
\begin{equation}
 \begin{array}{l}
 {{\mathcal{T}}_1=\frac{1}{8}\int dx_1\,dx_2\,dx_3\,\theta(x_1)\,H^{\mu_2}_{1,\nu_2}(x_2)H^{\mu_3}_{1,\nu_3}(x_3)\Big\{}\\[8pt]
 {\sum\limits_{i=1}^{f}\,c_i\,m_i\,
 [\xi^{(i)\alpha}\xi^{(i)}_{\alpha}(x_1)
 +\bar{\xi}^{(i)}_{\dot{\alpha}}\bar{\xi}^{(i)\,\dot{\alpha}}(x_1)]
 [j^{(i)\,\nu_2}_{R\,\mu_2}(x_2)-j^{(i)\,\nu_2}_{L\,\mu_2}(x_2)]
 [j^{(i)\,\nu_3}_{R\,\mu_3}(x_3)-j^{(i)\,\nu_3}_{L\,\mu_3}(x_3)]\rangle_{0}^{(c)}\Big]\Big\},}
 \end{array}
 \label{Tone}
 \end{equation}
 Here the left and right currents are defined as
 \bea
 &j^{(i)}_R\,_\m^\n(x)\equiv \overline{\xi}^{(i)}_{\dot{\a}}(x)\overline{\s}^{\dot{\a}\a \n}\pd_\m \xi_\a^{(i)}(x)\nonumber\\
 &j^{(i)}_L\,_\m^\n(x)\equiv  \pd_\m\overline{\xi}^{(i)}_{\dot{\a}}(x)\overline{\s}^{\dot{\a}\a \n}\xi_\a^{(i)}(x)
 \eea
\begin{equation}
\begin{array}{l}
{{\mathcal{T}}_2=\frac{1}{8}\,i\,\int dx_1\,dx_2\,dx_3\,\theta(x_1)\,h^{\mu_2}_{\mu_2}(x_2)H^{\mu_3}_{1,\nu_3}(x_3)\Big\{}\\[8pt]
{\sum\limits_{i=1}^{f}\,c_i\,m_i^2\,
[\xi^{(i)\alpha}\xi^{(i)}_{\alpha}(x_1)
+\bar{\xi}^{(i)}_{\dot{\alpha}}\bar{\xi}^{(i)\,\dot{\alpha}}(x_1)]
[\xi^{(i)\alpha}\xi^{(i)}_{\alpha}(x_2)
+\bar{\xi}^{(i)}_{\dot{\alpha}}\bar{\xi}^{(i)\,\dot{\alpha}}(x_2)]
[j^{(i)\,\nu_3}_{R\,\mu_3}(x_3)-j^{(i)\,\nu_3}_{L\,\mu_3}(x_3)]\rangle_{0}^{(c)}\Big]\Big\}}
\end{array}
\label{Ttwo}
\end{equation}
and
\begin{equation}
\begin{array}{l}
{{\mathcal{T}}_3=-\frac{1}{32}\int dx_1\,dx_2\,dx_3\,\theta(x_1)\,h^{\mu_2}_{\mu_2}(x_2)h^{\mu_3}_{\mu_3}(x_3)\Big\{}\\[8pt]
{\sum\limits_{i=1}^{f}\,c_i\,m_i^3\,
[\xi^{(i)\alpha}\xi^{(i)}_{\alpha}(x_1)
+\bar{\xi}^{(i)}_{\dot{\alpha}}\bar{\xi}^{(i)\,\dot{\alpha}}(x_1)]
[\xi^{(i)\alpha}\xi^{(i)}_{\alpha}(x_2)
+\bar{\xi}^{(i)}_{\dot{\alpha}}\bar{\xi}^{(i)\,\dot{\alpha}}(x_2)]
[\xi^{(i)\alpha}\xi^{(i)}_{\alpha}(x_3)
+\bar{\xi}^{(i)}_{\dot{\alpha}}\bar{\xi}^{(i)\,\dot{\alpha}}(x_3)]\Big\}.}\\[12pt]
\end{array}
\label{Tthree}
\end{equation}

\subsubsection{No Parity-odd Weyl anomaly from ${\mathcal{T}}_1$.}

Let us express ${\mathcal{T}}_1$ in (\ref{Tone}) in terms of the three different products of the  left and right currents and their symmetrizations
\begin{equation*}
{\mathcal{T}}_1=\frac{1}{8}\,\sum\limits_{m=1}^6\,{\mathcal{T}}_{1m},
\end{equation*}
 where
\begin{equation*}
\begin{array}{l}
{{\mathcal{T}}_{11}=\int dx_1\,dx_2\,dx_3\,\theta(x_1)\,H^{\mu_2}_{1,\nu_2}(x_2)H^{\mu_3}_{1,\nu_3}(x_3)\Big\{
\sum\limits_{i=1}^{f}\,c_i\,m_i\langle\xi^{(i)\alpha}\xi^{(i)}_{\alpha}(x_1)j^{(i)\,\nu_2}_{R\,\mu_2}(x_2)
j^{(i)\,\nu_3}_{R\,\mu_3}(x_3)\rangle_{0}^{(c)}\Big]\Big\},}\\[12pt]
{{\mathcal{T}}_{12}=\int dx_1\,dx_2\,dx_3\,\theta(x_1)\,H^{\mu_2}_{1,\nu_2}(x_2)H^{\mu_3}_{1,\nu_3}(x_3)\Big\{
\sum\limits_{i=1}^{f}\,c_i\,m_i\,\langle\bar{\xi}^{(i)}_{\dot{\alpha}}\bar{\xi}^{(i)\,\dot{\alpha}}(x_1)j^{(i)\,\nu_2}_{R\,\mu_2}(x_2)
 j^{(i)\,\nu_3}_{R\,\mu_3}(x_3)\rangle_{0}^{(c)}\Big]\Big\},}\\[12pt]

{{\mathcal{T}}_{13}=\int dx_1\,dx_2\,dx_3\,\theta(x_1)\,H^{\mu_2}_{1,\nu_2}(x_2)H^{\mu_3}_{1,\nu_3}(x_3)\Big\{
\sum\limits_{i=1}^{f}\,c_i\,m_i\langle\xi^{(i)\alpha}\xi^{(i)}_{\alpha}(x_1)j^{(i)\,\nu_2}_{L\,\mu_2}(x_2)
j^{(i)\,\nu_3}_{L\,\mu_3}(x_3)\rangle_{0}^{(c)}\Big]\Big\},}\\[12pt]
{{\mathcal{T}}_{14}=\int dx_1\,dx_2\,dx_3\,\theta(x_1)\,H^{\mu_2}_{1,\nu_2}(x_2)H^{\mu_3}_{1,\nu_3}(x_3)\Big\{
\sum\limits_{i=1}^{f}\,c_i\,m_i\,\langle\bar{\xi}^{(i)}_{\dot{\alpha}}\bar{\xi}^{(i)\,\dot{\alpha}}(x_1)j^{(i)\,\nu_2}_{L\,\mu_2}(x_2)
 j^{(i)\,\nu_3}_{L\,\mu_3}(x_3)\rangle_{0}^{(c)}\Big]\Big\},}\\[12pt]

{{\mathcal{T}}_{15}=-2\int dx_1\,dx_2\,dx_3\,\theta(x_1)\,H^{\mu_2}_{1,\nu_2}(x_2)H^{\mu_3}_{1,\nu_3}(x_3)\Big\{
\sum\limits_{i=1}^{f}\,c_i\,m_i\langle\xi^{(i)\alpha}\xi^{(i)}_{\alpha}(x_1)j^{(i)\,\nu_2}_{R\,\mu_2}(x_2)
j^{(i)\,\nu_3}_{L\,\mu_3}(x_3)\rangle_{0}^{(c)}\Big]\Big\},}\\[12pt]
{{\mathcal{T}}_{16}=-2\int dx_1\,dx_2\,dx_3\,\theta(x_1)\,H^{\mu_2}_{1,\nu_2}(x_2)H^{\mu_3}_{1,\nu_3}(x_3)\Big\{
\sum\limits_{i=1}^{f}\,c_i\,m_i\,\langle\bar{\xi}^{(i)}_{\dot{\alpha}}\bar{\xi}^{(i)\,\dot{\alpha}}(x_1)j^{(i)\,\nu_2}_{L\,\mu_2}(x_2)
 j^{(i)\,\nu_3}_{R\,\mu_3}(x_3)\rangle_{0}^{(c)}\Big]\Big\}. }
\end{array}
\end{equation*}
Recall that  ${\mathcal{T}}_1$ is symmetric under the exchange of $(\mu_2,\nu_2,x_2)$ and  $(\mu_3,\nu_3,x_3)$.

The application of Wick's theorem readily leads to the following results for the parity-odd contributions of ${\mathcal{T}}_{1m}$:
\begin{equation*}
\begin{array}{l}
{{\mathcal{T}}_{11}[odd]
= 4\int\prod\limits_{i=1}^{3}\dpi(2\pi)^4\,\delta(p_1+p_2+p_3)\,\theta(p_1)\,H^{\mu_2}_{1,\nu_2}(p_2)H^{\mu_3}_{1,\nu_3}(p_3)\Big\{}\\
{\idq\big[\sum\limits_{i=1}^{f}c_im_i^2\cfrac{\epsilon^{\nu_4\nu_5\nu_2\nu_3}[
(q+p_3)_{\mu_2}(q+p_3)_{\mu_3}(p_2+p_3)_{\nu_4} q_{\nu_5}-2(q+p_2+p_3)_{\mu_2}(q+p_3)_{\mu_3} p_{3\nu_4}q_{\nu_5}] }
{(q^2-m_i^2)((q+p_3)^2-m_i^2)((q+p_2+p_3)^2-m_i^2)}\big]\Big\}}\\[12pt]

{{\mathcal{T}}_{12}
[odd]=4\int\prod\limits_{i=1}^{3}\dpi(2\pi)^4\,\delta(p_1+p_2+p_3)\,\theta(p_1)\,H^{\mu_2}_{1,\nu_2}(p_2)H^{\mu_3}_{1,\nu_3}(p_3)\Big\{}\\
{\idq\big[\sum\limits_{i=1}^{f}c_im_i^2\cfrac{\epsilon^{\nu_4\nu_5\nu_2\nu_3}[
2(q+p_2+p_3)_{\mu_2}(q+p_3)_{\mu_3} p_{2\nu_4}q_{\nu_5}-(q+p_2+p_3)_{\mu_2}q_{\mu_3}(p_2+p_3)_{\nu_4}q_{\nu_5}]
}{(q^2-m_i^2)((q+p_3)^2-m_i^2)((q+p_2+p_3)^2-m_i^2)}\big]\Big\}}\\[12pt]

{{\mathcal{T}}_{13}[odd]=-{\mathcal{T}}_{12}[odd],}\\[12pt]
{{\mathcal{T}}_{14}[odd]=-{\mathcal{T}}_{11}[odd],}\\[12pt]

{{\mathcal{T}}_{15}[odd]=-8\int\prod\limits_{i=1}^{3}\dpi(2\pi)^4\,\delta(p_1+p_2+p_3)\,\theta(p_1)\,H^{\mu_2}_{1,\nu_2}(p_2)H^{\mu_3}_{1,\nu_3}(p_3)\Big\{}\\
 {\idq\big[\sum\limits_{i=1}^{f}c_im_i^2\cfrac{\epsilon^{\nu_4\nu_5\nu_2\nu_3}[
p_{2\nu_4}p_{3\nu_5}(q+p_3)_{\mu_2}(q+p_3)_{\mu_3}+ p_{2\nu_4}q_{\nu_5}(q+p_3)_{\mu_2}p_{3\mu_3}+p_{3\nu_4}q_{\nu_5}p_{2\mu_2}q_{\mu_3}]
 }{(q^2-m_i^2)((q+p_3)^2-m_i^2)((q+p_2+p_3)^2-m_i^2)}\big]\Big\},}\\[12pt]

{{\mathcal{T}}_{16}[odd]=-{\mathcal{T}}_{15}[odd].}
\end{array}
\end{equation*}

It follows that,

\begin{equation*}
 {\mathcal{T}}_{11}[odd]+ {\mathcal{T}}_{14}[odd]=0,\quad  {\mathcal{T}}_{12}[odd]+ {\mathcal{T}}_{13}[odd]=0,\quad   {\mathcal{T}}_{15}[odd]+{\mathcal{T}}_{16}[odd]=0,
\end{equation*}
so that no parity-odd contribution to the Weyl anomaly comes from ${\mathcal{T}}_1$.

\subsubsection{No Parity-odd Weyl anomaly from ${\mathcal{T}}_2$.}

 The only contributions to ${\mathcal{T}}_2$  in (\ref{Ttwo}) which may give rise to a trace over four Pauli matrices,
 and thus an odd parity contribution are:
 \begin{equation}
 \begin{array}{l}
 {{\mathcal{T}}_{21}=\big(\frac{1}{8}\,i\,\big)\int dx_1\,dx_2\,dx_3\,
 [\theta(x_1)\,h^{\mu_2}_{\mu_2}(x_2)+\theta(x_2)\,h^{\mu_1}_{\mu_1}(x_1)]H^{\mu_3}_{1,\nu_3}(x_3) \Big\{}\\[8pt]
 {\phantom{{\cal C}_{51}=\big(-\frac{i}{8}\big)}
 \sum\limits_{i=1}^{f}\,c_i\,m_i^2\,
 \Big[\langle\xi^{(i)\alpha_1}(x_1)\xi^{(i)}_{\alpha_1}(x_1)\bar{\xi}^{(i)}_{\dot{\alpha}_2}(x_2)\bar{\xi}^{(i)\,\dot{\alpha}_2}(x_2)\,
 j^{(i)\,\nu_3}_{R\,\mu_3}(x_3)\rangle_{0}^{(c)}\Big]\Big\},}\\[12pt]

 {{\mathcal{T}}_{22}=\big(-\frac{1}{8}\,i\,\big)\int dx_1\,dx_2\,dx_3\,
 [\theta(x_1)\,h^{\mu_2}_{\mu_2}(x_2)+\theta(x_2)\,h^{\mu_1}_{\mu_1}(x_1)]H^{\mu_3}_{1,\nu_3}(x_3)
 \Big\{}\\[8pt]
 {\phantom{{\cal C}_{51}=\big(-\frac{7}{4}\big)}
 \sum\limits_{i=1}^{f}\,c_i\,m_i^2\,
 \Big[\langle\bar{\xi}^{(i)}_{\dot{\alpha}_1}(x_1)\bar{\xi}^{(i)\,\dot{\alpha}_1}(x_1)\xi^{(i)\alpha_2}(x_2)\xi^{(i)}_{\alpha_2}(x_2)
 j^{(i)\,\nu_3}_{L\,\mu_3}(x_3)\rangle_{0}^{(c)}\Big]\Big\},}
 \end{array}
 \label{T21}
 \end{equation}

 A tedious computation  shows that the odd parity contribution to ${\mathcal{T}}_{21}$ reads
 \begin{equation}
 \begin{array}{l}
 {{\mathcal{T}}_{21}[odd]=-\int\prod\limits_{i=1}^{3}\dpi(2\pi)^4\,\delta(p_1+p_2+p_3)[\theta(p_1)\,h^{\mu_2}_{\mu_2}(p_2)+
 \theta(p_2)\,h^{\mu_1}_{\mu_1}(p_1)]
 H^{\mu_3}_{1,\nu_3}(p_3)\Big\{}\\
 {\idq\big[\sum\limits_{i=1}^{f}c_im_i^2\cfrac{\epsilon^{\nu_1\nu_2\nu_3\nu_4}\,
 p_{2\nu_1} p_{3\nu_2}(q+p_3)_{\mu_3} q_{\nu_4} }
 {(q^2-m_i^2)((q+p_3)^2-m_i^2)((q+p_2+p_3)^2-m_i^2)}\big]\Big\},}
 \end{array}
 \label{T21oddres}
 \end{equation}
 whereas the odd parity contribution to ${\mathcal{T}}_{22}[odd]$ is
 \begin{equation}
 {\mathcal{T}}_{22}[odd]=-{\mathcal{T}}_{21}[odd].
 \label{T22oddres}
 \end{equation}

 Putting it all together we conclude that ${\mathcal{T}}_{2}$ carries no parity-odd contributions.

 \subsubsection{No parity-odd anomaly from ${\mathcal{T}}_{3}$.}

 ${\mathcal{T}}_{3}$ in  (\ref{Tthree}) never yields parity-odd contributions since the maximum number of Pauli Matrices to trace over
  is two.

\subsection{No parity-odd anomaly from bubble-type contributions.}
There are also terms that would give rise under contractions to {\em bubble}  type Feynman diagrams. These terms carry one variation of the PV mass term and a kinetic term to second order.
Here, we shall show that no parity-odd contribution to the Weyl anomaly will arise from either
\begin{equation}
{\mathcal{B}}_{1} = \sum\limits_{i=1}^f\,c_i\,\langle \delta^{(0)}_\theta S^{(i)}_{m}  S^{(i),2}_{kin}\rangle_0^{(c)}
\label{Beeone}
\end{equation}
 or else
\begin{equation}
{\mathcal{B}}_{2} = \sum\limits_{i=1}^f\,c_i\,\langle \delta^{(0)}_\theta S^{(i)}_{m}  S^{(i),2}_{spin}\rangle_0^{(c)}
\label{Beetwo}
\end{equation}
or again,
\begin{equation}
{\mathcal{B}}_{3} = \sum\limits_{i=1}^f\,c_i\,\langle \delta^{(1)}_\theta S^{(i)}_{m}  S^{(i),1}_{kin}\rangle_0^{(c)}
\label{Beethree}
\end{equation}
in (\ref{Anomalytwo}).

We can now consider separately terms independent of the PV masses and those which are not.
\begin{equation*}
 {\mathcal{B}}_{1}={\mathcal{B}}_{11} +{\mathcal{B}}_{12}, \end{equation*}
with
\begin{equation*}
\begin{array}{l}
{{\mathcal{B}}_{11}=-\frac{1}{4}\,i\,\int dx_1 dx_2\; \theta(x_1)H_{2\nu_2}^{\mu_2}(x_2)\big\{}\\[4pt]
{\phantom{{\cal B}_{21}=-\frac{1}{2}\,\int dx_1 }
\sum\limits_{i=0}^f\,c_i\,m_i\,
\langle [\xi^{(i)\alpha_1}(x_1)\xi^{(i)}_{\alpha_1}(x_1)+\bar{\xi}^{(i)}_{\dot{\alpha}_1}(x_1)\bar{\xi}^{(i)\dot{\alpha}_1}(x_1)]
[j_{R\mu_2}^{(i)\nu_2}(x_2)-j_{L\mu_2}^{(i)\nu_2}(x_2)]\rangle^{(c)}_0\big\},}\\[12pt]
{{\mathcal{B}}_{12}=\frac{1}{8}\,\int dx_1 dx_2\;\theta(x_1)H_{2}(x_2)
\big\{}\\[4pt]
{
\sum\limits_{i=0}^f\,c_i\,m_i^2\,
\langle [\xi^{(i)\alpha_1}(x_1)\xi^{(i)}_{\alpha_1}(x_1)+\bar{\xi}^{(i)}_{\dot{\alpha}_1}(x_1)\bar{\xi}^{(i)\dot{\alpha}_1}(x_1)]
[\xi^{(i)\alpha_2}(x_2)\xi^{(i)}_{\alpha_2}(x_2)+
\bar{\xi}^{(i)}_{\dot{\alpha}_2}(x_2)\bar{\xi}^{(i)\dot{\alpha}_2}(x_2)] \rangle^{(c)}_0\big\}.}\\[12pt]
\end{array}
\end{equation*}
 Now, ${\mathcal{B}}_{11}$ and ${\mathcal{B}}_{12}$ cannot give rise to parity-odd contributions since, at most, the trace of two Pauli matrices occurs.
 Hence,  ${\mathcal{B}}_{1}$ in (\ref{Beeone}) yields no parity-odd contribution.

Let us move on and  analyze ${\mathcal{B}}_{2}$ in (\ref{Beetwo}). Let us express  ${\mathcal{B}}_{2}$ as follows
\begin{equation*}
   {\mathcal{B}}_{2}={\mathcal{B}}_{21}+ {\mathcal{B}}_{22},
\end{equation*}
where
\begin{equation}
\begin{array}{l}
{{\mathcal B}_{21}=-\frac{1}{2}\int dx_1 dx_2\;\theta(x_1) H_{\mu\nu\rho}(x_2)\big\{}\\[4pt]
{\phantom{{\cal B}_{21}=-\frac{1}{2}\,\int dx_1 dx_2[\delta^{(0)}_\theta H_{1\nu_1}}
\epsilon^{\mu\nu\rho\nu_2}\sum\limits_{i=0}^f\,c_i\,m_i\,
\langle \xi^{(i)\alpha_1}(x_1)\xi^{(i)}_{\alpha_1}(x_1)\,
\bar{\xi}^{(i)}_{\dot{\alpha}_2}(x_2)\bsigma^{\dot{\alpha}_2\alpha_2}_{\nu_2}\xi^{(i)}_{\alpha_2}(x_2)\rangle^{(c)}_0\big\},}\\[12pt]

{{\mathcal B}_{22}=-\frac{1}{2}\int dx_1 dx_2\;\theta(x_1) H_{\mu\nu\rho}(x_2)\big\{}\\[4pt]
{\phantom{{\cal B}_{21}=-\frac{1}{2}\,\int dx_1 dx_2[\delta^{(0)}_\theta H_{1\nu_1}}
\epsilon^{\mu\nu\rho\nu_2}\sum\limits_{i=0}^f\,c_i\,m_i\,
\langle \bar{\xi}^{(i)}_{\dot{\alpha}_1}(x_1)\bar{\xi}^{(i)\dot{\alpha}_1}(x_1)
\bar{\xi}^{(i)}_{\dot{\alpha}_2}(x_2)\,\bsigma^{\dot{\alpha}_2\alpha_2}_{\nu_2}\xi^{(i)}_{\alpha_2}(x_2)\rangle^{(c)}_0\big\},}\\[12pt]
{H_{\mu\nu\rho}(x)=\frac{1}{16}\,h^{\lambda}_\mu(x)\partial_\nu h_{\rho\lambda}(x).}
\end{array}
\end{equation}

A little computation yields
  \begin{equation*}
  \begin{array}{l}
  {{\mathcal B}_{21}=-2\int\prod\limits_{i=1}^{2}\dpi(2\pi)^4\,\delta(p_1+p_2) \theta(p_1)
  H_{\mu\nu\rho}(p_2)\Big\{
  \idq\,\Big[\sum\limits_{i=1}^{f}\,c_i\,m_i^2\cfrac{\epsilon^{\mu\nu\rho\lambda}(q+p_1)_{\lambda}}
  {(q^2-m_i^2)((q+p_1)^2-m_i^2)}\Big]\Big\}}\\[12pt]
  {{\mathcal B}_{22}=-{\mathcal B}_{21},}
  \end{array}
  \end{equation*}
  so that
  \begin{equation*}
  {\mathcal B}_{21}+{\mathcal B}_{22}=0.
  \end{equation*}
Hence, we conclude that  ${\mathcal B}_{2}$ in (\ref{Beetwo}) gives rise to no parity-odd anomalous term.
Finally,
 \begin{equation*}
  {\mathcal{B}}_{3}={\mathcal{B}}_{31} +{\mathcal{B}}_{32},
 \end{equation*}
 with
 \begin{equation*}
 \begin{array}{l}
 {{\mathcal{B}}_{31}=-\frac{1}{8}\,i\,\int dx_1 dx_2\; \theta(x_1) h^{\mu_1}_{\mu_1}(x_1)\,H_{1\nu_2}^{\mu_2}(x_2)\big\{}\\[4pt]
 {\phantom{{\cal B}_{21}=-\frac{1}{2}\,\int dx_1 }
 \sum\limits_{i=0}^f\,c_i\,m_i\,
 \langle [\xi^{(i)\alpha_1}(x_1)\xi^{(i)}_{\alpha_1}(x_1)+\bar{\xi}^{(i)}_{\dot{\alpha}_1}(x_1)\bar{\xi}^{(i)\dot{\alpha}_1}(x_1)]
 [j_{R\mu_2}^{(i)\nu_2}(x_2)-j_{L\mu_2}^{(i)\nu_2}(x_2)]\rangle^{(c)}_0\big\},}\\[12pt]

 {{\mathcal{B}}_{32}=\frac{1}{16}\,\int dx_1 dx_2\;\theta(x_1)h^{\m_1}_{\m_1}(x_1)\,h^{\mu_2}_{\mu_2}(x_2)
 \big\{}\\[4pt]
 {
 \sum\limits_{i=0}^f\,c_i\,m_i^2\,
 \langle [\xi^{(i)\alpha_1}(x_1)\xi^{(i)}_{\alpha_1}(x_1)+\bar{\xi}^{(i)}_{\dot{\alpha}_1}(x_1)\bar{\xi}^{(i)\dot{\alpha}_1}(x_1)]
 [\xi^{(i)\alpha_2}(x_2)\xi^{(i)}_{\alpha_2}(x_2)+
 \bar{\xi}^{(i)}_{\dot{\alpha}_2}(x_2)\bar{\xi}^{(i)\dot{\alpha}_2}(x_2)] \rangle^{(c)}_0\big\}.}\\[12pt]
 \end{array}
 \end{equation*}
 But ${\mathcal{B}}_{31}$ and ${\mathcal{B}}_{32}$ cannot give rise to parity-odd contributions, since the maximum number of Pauli matrices involved is two. Hence,  ${\mathcal{B}}_{3}$ in (\ref{Beethree}) produces no parity-odd contribution.

 \subsection{No parity-odd anomaly from tadpole-type contributions.}
The terms in the second order variation of the PV mass terms would give rise to tadpoles upon Wicks contractions.
The tadpole-type contributions to the anomaly, if any, comes from
\begin{equation*}
\sum\limits_{i=1}^f\,c_i\,\langle \delta^{(2)}_\theta S^{(i)}_{m}\rangle_0^{(c)}.
 \end{equation*}
Since this contribution involves no Pauli matrices, there is no parity-odd term coming from it.

\section{Conclusions.}

When computing the Weyl anomaly in perturbation theory one has to deal with the follow type of vacuum expectation values
\begin{equation}
\langle 0|\delta S_{int} S_{int}\cdots S_{int}|0\rangle=\int dx\,dx_1\cdots dx_n\,\langle 0|\delta{\cal L}_{int}(x){\cal L}_{int}(x_1)\cdots {\cal L}_{int}(x_n)|0\rangle
\label{VEV}
\end{equation}
where $|0\rangle$ is the Fock vacuum of the theory and $S_{int}$ the part of the action which contains the interaction and $\delta S_{int}$ is its symmetry variation. Obviously, we assume that $S_{int}$ is hermitian and so is $\delta S_{int}$.
Formally, the vacuum expectation value in question is real;
but this is not a mathematically rigorous statement, for, in general, the correlation functions on the right hand side  of (\ref{VEV}) contain UV divergences, which
must be regularized. If there is a regularization method that explicitly preserves the formal reality of  correlation functions under scrutiny,
it is clear that no parity-odd contribution to the Weyl anomaly should show up. In this paper we have put forward such a regularization method --a Pauli-Villars regularization-- based on the use of a  real Lagrangian (not just a real Lagrangian modulo a total derivative) for each Pauli-Villars field.
Notice, however, that the regularized correlation functions are well-defined once they are represented in momentum space, and once the loop contributions coming from the
physical fields and the Pauli-Villars fields have been collected  as a linear combination inside the loop integrations. Hence, to show that the correlation functions which may give rise to a parity-odd contribution to the anomaly do cancel, we have checked that for each contribution of this type there is always a contribution with opposite sign.
Finally, we believe that the formal reality of
the correlation functions should be preserved in the regularized theory since it comes from the reality of the classical action and should be taken as a
 fundamental property of the quantum theory.
 \par
 Let us end with a comment. There is no known topological interpretation of the Weyl anomaly; in particular no known relationship with any index theorem.
 In this respect it is interesting to highlight the work  \cite{Bittleston} where the exact form of the one-loop beta function of QCD is recovered from an index theorem
on twistor space .

\section{Acknowledgements}
We acknowkledge stimulating discussions with Peter van Nieuwenhuizen and Zohar Komargodski. We also acknowledge partial financial support by the
 Spanish MINECO through the Centro de excelencia Severo Ochoa Program  under Grant CEX2020-001007-S  funded by MCIN/AEI/10.13039/501100011033.
We also acknowledge partial financial support by the Spanish Research Agency (Agencia Estatal de Investigaci\'on) through the grant PID2022-137127NB-I00 funded by MCIN/AEI/10.13039/501100011033/ FEDER, UE.
All authors acknowledge the European Union's Horizon 2020 research and innovation programme under the Marie Sklodowska-Curie grant agreement No 860881-HIDDeN and also byGrant PID2019-108892RB-I00 funded by MCIN/AEI/ 10.13039/501100011033 and by ``ERDF A way of making Europe''. CPM's research work has been financially supported in part by the Spanish Ministry of
Science, Innovation and Universities through grant PID2023-149834NB-I00.

\newpage

\appendix

\section{Some computational details.}

As an example of the standard techniques used to obtain the results displayed in this paper, we shall give here some details of the computations to be done to reproduce the results in subsection 2.1.2.

To calculate ${\mathcal{T}}_{21}$ in (\ref{T21}), one needs to work out the correlation function
\begin{equation}
\langle\xi^{(i)\alpha_1}(x_1)\xi^{(i)}_{\alpha_1}(x_1)\bar{\xi}^{(i)}_{\dot{\alpha}_2}(x_2)\bar{\xi}^{(i)\,\dot{\alpha}_2}(x_2)\,
\overline{\xi}^{(i)}_{\dot{\a_3}}(x_3)\overline{\s}^{\dot{\a_3}\a_3 \n_3}\pd_{\m_3} \xi_{\a_3}^{(i)}(x_3)\rangle_{0}^{(c)}.
\label{correlator}
\end{equation}
A trivial application of Wick's theorem leads to the conclusion that the previous Green function is equal to
\begin{equation}
\begin{array}{l}
{4\,\epsilon^{\alpha_1\beta_1}\,\epsilon^{\dot{\alpha}_2\dot{\beta}_2}\,\overline{\s}^{\dot{\a_3}\a_3 \n_3}\Big[
\langle\xi^{(i)}_{\alpha_1}(x_1)\bar{\xi}^{(i)}_{\dot{\alpha}_2}(x_2)\rangle_{0}\,
\langle\bar{\xi}^{(i)}_{\dot{\beta}_2}(x_2)\pd_{\m_3}\xi^{(i)}_{\alpha_3}(x_3)\rangle_0\,
\langle\bar{\xi}^{(i)}_{\dot{\alpha}_3}(x_3)\xi^{(i)}_{\beta_1}(x_1)\rangle_0}\\[4pt]
{\phantom{4\,\epsilon^{\alpha_1\beta_1}\,\epsilon^{\dot{\alpha}_2\dot{\beta}_2}\,\bar{\s}^{\dot{\a_3}\a_3 \n_3}\Big[}
-\langle\xi^{(i)}_{\alpha_1}(x_1)\bar{\xi}^{(i)}_{\dot{\alpha}_2}(x_2)\rangle_{0}\,\langle\bar{\xi}^{(i)}_{\dot{\beta}_2}(x_2)\bar{\xi}^{(i)}_{\dot{\alpha}_3}(x_3)\rangle_0\,
\langle\pd_{\m_3}\xi^{(i)}_{\alpha_3}(x_3)\xi^{(i)}_{\beta_1}(x_1)\rangle_0\Big].}
\end{array}
\label{Wickexp}
\end{equation}
Now, the free propagators --see \cite{Dreiner}-- in the previous equation read
\begin{equation}
\begin{array}{l}
{\langle\xi^{(i)}_{\alpha}(x)\bar{\xi}^{(i)}_{\dot{\alpha}}(y)\rangle_{0}=\idp\, p\sigma_{\alpha\dot{\alpha}}\,G_i(p)\,e^{-ip(x-y)},}\\[4pt]
{\langle{\bar\xi}^{(i)}_{\dot{\alpha}}(x)\partial_\mu\xi^{(i)}_{\alpha}(y)\rangle_{0}=\idp\,(ip_\mu)\,  p\sigma_{\alpha\dot{\alpha}}\,G_i(p)\,e^{-ip(x-y)},}\\[4pt]
{\langle\xi^{(i)}_{\alpha}(x)\xi^{(i)}_{\beta}(y)\rangle_{0}=\idp\, m_i\, \epsilon_{\beta\alpha}\,G_i(p)\,e^{-ip(x-y)},}\\[4pt]
{\langle{\bar\xi}^{(i)}_{\dot{\alpha}}(x)\bar{\xi}^{(i)}_{\dot{\beta}}(y)\rangle_{0}=\idp\, m_i\, \epsilon_{\dot{\alpha}\dot{\beta}}\,G_i(p)\,e^{-ip(x-y)},}\\[4pt]
{\langle\partial_\mu\xi_{\alpha}^{(i)}(x)\xi^{(i)}_{\beta}(y)\rangle_{0}=\idp\, m_i\, (-i p_\mu)\epsilon_{\beta\alpha}\,G_i(p)\,e^{-ip(x-y)}},\\[4pt]
{G_i(p)=\frac{i}{p^2-m_i^2}.}
\end{array}
\label{freeprop}
\end{equation}

The substitution of the free propagators in (\ref{freeprop}) in equation (\ref{Wickexp}), and some algebra, yields  the correlation function in (\ref{correlator}) in terms of its Fourier transform. This expression for the Green function in (\ref{freeprop}) substituted in (\ref{T21}) yields (\ref{T21oddres}). Notice that, on the one hand,
the product of three free propagators in  the second line of (\ref{Wickexp}) does not contribute to the odd parity part of ${\mathcal{T}}_{21}$, for it leads to the trace of the product of only two Pauli matrices. On the other hand, the computation of the contribution to ${\mathcal{T}}_{21}$ coming from the first line of (\ref{Wickexp}) demands to work out the trace over the product of four Pauli matrices, which yields a term involving the $\epsilon^{\mu\nu\rho\sigma}$  pseudo-tensor.

Finally, equation (\ref{T22oddres}) can be obtained in a completely analogous way.



\begin{thebibliography}{99}
 \bibitem{Abdallah}
S.~Abdallah, S.~A.~Franchino-Vi{\~n}as and M.~B.~Fr{\"o}b,\\
``Trace anomaly for Weyl fermions using the Breitenlohner-Maison scheme for $\gamma_*$,''
JHEP \textbf{03} (2021), 271
doi:10.1007/JHEP03(2021)271
[arXiv:2101.11382 [hep-th]].\\
``Trace anomalies for Weyl fermions: too odd to be true?,''
J. Phys. Conf. Ser. \textbf{2531} (2023) no.1, 012004
doi:10.1088/1742-6596/2531/1/012004
[arXiv:2304.08939 [hep-th]].
 \bibitem{Bastianelli}
F.~Bastianelli and R.~Martelli,
``On the trace anomaly of a Weyl fermion,''
JHEP \textbf{11} (2016), 178
doi:10.1007/JHEP11(2016)178
[arXiv:1610.02304 [hep-th]].\\
F.~Bastianelli and M.~Broccoli,
``Axial gravity and anomalies of fermions,''
Eur. Phys. J. C \textbf{80} (2020) no.3, 276
doi:10.1140/epjc/s10052-020-7782-4
[arXiv:1911.02271 [hep-th]].\\
F.~Bastianelli and M.~Broccoli,
``On the trace anomaly of a Weyl fermion in a gauge background,''
Eur. Phys. J. C \textbf{79} (2019) no.4, 292
doi:10.1140/epjc/s10052-019-6799-z
[arXiv:1808.03489 [hep-th]].\\
F.~Bastianelli and M.~Broccoli,
``Weyl fermions in a non-abelian gauge background and trace anomalies,''
JHEP \textbf{10} (2019), 241
doi:10.1007/JHEP10(2019)241
[arXiv:1908.03750 [hep-th]].\\
F.~Bastianelli and F.~Comberiati,
``Path integral calculation of heat kernel traces with first order operator insertions,''
Nucl. Phys. B \textbf{960} (2020), 115183
doi:10.1016/j.nuclphysb.2020.115183
[arXiv:2005.08737 [hep-th]].
\bibitem{Bittleston}
R.~Bittleston and K.~Costello,
``The One-Loop QCD $\beta$-Function as an Index,''
[arXiv:2510.26764 [hep-th]].
\bibitem{Bonora}
L.~Bonora, S.~Giaccari and B.~Lima de Souza,
``Trace anomalies in chiral theories revisited,''
JHEP \textbf{07} (2014), 117
doi:10.1007/JHEP07(2014)117
[arXiv:1403.2606 [hep-th]].\\
L.~Bonora and R.~Soldati,
``On the trace anomaly for Weyl fermions,''
[arXiv:1909.11991 [hep-th]].\\
``Anomaly footprints in SM+Gravity,''
[arXiv:2510.25217 [hep-th]].
\bibitem{Book}
L.~Bonora,
``Fermions and Anomalies in Quantum Field Theories,''
Springer, 2023,
ISBN 978-3-031-21927-6, 978-3-031-21928-3
doi:10.1007/978-3-031-21928-3
\bibitem{Christensen}
S.~M.~Christensen and M.~J.~Duff,
``Axial and Conformal Anomalies for Arbitrary Spin in Gravity and Supergravity,''
Phys. Lett. B \textbf{76} (1978), 571
doi:10.1016/0370-2693(78)90857-2\\
``New Gravitational Index Theorems and Supertheorems,''
Nucl. Phys. B \textbf{154} (1979), 301-342
doi:10.1016/0550-3213(79)90516-9


\bibitem{Dreiner}
H.~K.~Dreiner, H.~E.~Haber and S.~P.~Martin,
``Two-component spinor techniques and Feynman rules for quantum field theory and supersymmetry,''
Phys. Rept. \textbf{494} (2010), 1-196
doi:10.1016/j.physrep.2010.05.002
[arXiv:0812.1594 [hep-ph]].
\bibitem{Eguchi}
T.~Eguchi, P.~B.~Gilkey and A.~J.~Hanson,
``Gravitation, Gauge Theories and Differential Geometry,''
Phys. Rept. \textbf{66} (1980), 213
doi:10.1016/0370-1573(80)90130-1
\bibitem{Faddeev}
L.~D.~Faddeev and A.~A.~Slavnov,
``GAUGE FIELDS. INTRODUCTION TO QUANTUM THEORY,''
Front. Phys. \textbf{50} (1980), 1-232
\bibitem{Frob}
M.~B.~Fr{\"o}b and J.~Zahn,
``Trace anomaly for chiral fermions via Hadamard subtraction,''
JHEP \textbf{10} (2019), 223
doi:10.1007/JHEP10(2019)223
[arXiv:1904.10982 [hep-th]].\\
``The Weyl anomaly in interacting quantum field theory on curved spacetimes,''
doi:10.1007/s00023-025-01635-2
[arXiv:2504.17854 [hep-th]].

\bibitem{Fujikawa}
K.~Fujikawa,
`Comment on Chiral and Conformal Anomalies,''
Phys. Rev. Lett. \textbf{44} (1980), 1733
doi:10.1103/PhysRevLett.44.1733
\bibitem{Gipson}
J.~M.~Gipson,
``Path Integral Derivation of the Chiral Anomalies of Dirac Fermions Coupled to Gauge and Gravitational Fields in Even Space-time Dimensions,''
Phys. Rev. D \textbf{29} (1984), 2989
doi:10.1103/PhysRevD.29.2989
\bibitem{Godazgar}
H.~Godazgar and H.~Nicolai,
``A rederivation of the conformal anomaly for spin-$\frac{1}{2}$,''
Class. Quant. Grav. \textbf{35} (2018) no.10, 105013
doi:10.1088/1361-6382/aaba97
[arXiv:1801.01728 [hep-th]].

\bibitem{Itzykson}
C.~Itzykson and J.~B.~Zuber,
``Quantum Field Theory,''
McGraw-Hill, 1980,
ISBN 978-0-486-44568-7

\bibitem{Larue}
R.~Larue, J.~Quevillon and R.~Zwicky,
``Gravity-gauge anomaly constraints on the energy-momentum tensor,''
JHEP \textbf{05} (2024), 307
doi:10.1007/JHEP05(2024)307
[arXiv:2312.13222 [hep-th]].\\
R.~Larue, J.~Quevillon and R.~Zwicky,
``Trace anomaly of weyl fermions via the path integral,''
JHEP \textbf{12} (2023), 064
doi:10.1007/JHEP12(2023)064
[arXiv:2309.08670 [hep-th]].
\bibitem{Leutwyler}
H.~Leutwyler and S.~Mallik,
``GRAVITATIONAL ANOMALIES,''
Z. Phys. C \textbf{33} (1986), 205
doi:10.1007/BF01411138
\bibitem{Liu}
C.~Y.~Liu,
``Investigation of Pontryagin trace anomaly using Pauli-Villars regularization,''
Nucl. Phys. B \textbf{980} (2022), 115840
doi:10.1016/j.nuclphysb.2022.115840
[arXiv:2202.13893 [hep-th]].
\bibitem{Okuyama}
K.~Okuyama and H.~Suzuki,
 ``Gauge invariant Pauli-Villars regularization for chiral fermion,''
Prog. Theor. Phys. \textbf{98} (1997), 463-484
doi:10.1143/PTP.98.463
[arXiv:hep-th/9603062 [hep-th]].
\bibitem{Stanev}
Y.~S.~Stanev,
``Correlation Functions of Conserved Currents in Four Dimensional Conformal Field Theory,''
Nucl. Phys. B \textbf{865} (2012), 200-215
doi:10.1016/j.nuclphysb.2012.07.027
[arXiv:1206.5639 [hep-th]].
\bibitem{Vilenkin}
A. Vilenkin,
``Pauli-Villars Regularization and Trace Anomalies."
Nuovo Cim. L. 44A, N. 3 1 April 1978
\end{thebibliography}
\end{document}